\def\ls{\lower4pt\hbox{${\buildrel < \over \sim}$}}
\def\gs{\lower4pt\hbox{${\buildrel > \over \sim}$}}

\documentclass[12pt, preprint]{aastex}

\slugcomment{Accepted for Publication in {\it The Astrophysical Journal}}

\shorttitle{On the VHE Detection of 3C279}
\shortauthors{B\"ottcher et al.}

\begin{document}

\title{Implications of the VHE Gamma-Ray Detection of the Quasar 3C279}

\author{M. B\"ottcher\altaffilmark{1} \and A. Reimer\altaffilmark{2}
\and A. P. Marscher\altaffilmark{3}
}

\altaffiltext{1}{Astrophysical Institute, Department of Physics and Astronomy, \\
Clippinger 339, Ohio University, Athens, OH 45701, USA}
\altaffiltext{2}{Stanford University, HEPL/KIPAC, Stanford, CA 94305, USA}
\altaffiltext{3}{Institute of Astrophysical Research, Boston University, \\
725 Commonwealth Avenue, Boston, MA 02215, USA}

\begin{abstract}
The MAGIC very-high-energy (VHE) $\gamma$-ray astronomy collaboration 
recently reported the detection of the quasar 3C279 at $> 100$~GeV 
$\gamma$-ray energies. Here we present simultaneous optical (BVRI) 
and X-ray ({\it RXTE PCA}) data from the day of the VHE detection and 
discuss the implications of the snap-shot spectral energy distribution 
for jet models of blazars. A one-zone synchrotron-self-Compton origin 
of the entire SED, including the VHE $\gamma$-ray emission is highly 
problematic as it would require an unrealistically low magnetic field.
The measured level of VHE emission could, in principle, be 
interpreted as Compton upscattering of external radiation (e.g., from 
the broad-line regions). However, such an interpretation would require 
either an unusually low magnetic field of $B \sim 0.03$~G, or (in order 
to achieve approximate equipartition between magnetic field at $B \sim 
0.25$~G and relativistic electrons) an unrealistically high Doppler 
factor of $\Gamma \sim 140$. In addition, such a model fails to reproduce 
the observed X-ray flux. Furthermore, both versions of leptonic
one-zone models produce intrinsic VHE $\gamma$-ray spectra steeper
than measured, even in the case of the lowest plausible extragalactic
$\gamma\gamma$ absorption.
We therefore
conclude that a simple one-zone, homogeneous leptonic jet model is
not able to plausibly reproduce the SED of 3C279 including the recently
detected VHE $\gamma$-ray emission. This as well as the lack of correlated
variability in the optical with the VHE $\gamma$-ray emission and the
substantial $\gamma\gamma$ opacity of the BLR radiation field to VHE
$\gamma$-rays suggests a multi-zone model in which the optical emission 
is produced in a different region than the VHE $\gamma$-ray 
emission. In particular, an SSC model with an emission region far outside
the BLR reproduces the simultaneous X-ray --- VHE $\gamma$-ray spectrum 
of 3C279. Alternatively, a hadronic model is capable of reproducing 
the observed SED of 3C279 reasonably well, both in scenarios in which
only the internal synchrotron field serves as targets for $p\gamma$
pion production, and with a substantial contribution from external
photons, e.g., from the BLR. However, either version of the hadronic
model requires a rather extreme jet power of 
up to
$L_j \sim 10^{49}$~erg~s$^{-1}$,
compared to a requirement of $L_j \sim 2 \times 10^{47}$~erg~s$^{-1}$ for
a multi-zone leptonic model.
\end{abstract}


\section{Introduction}

Flat-spectrum radio quasars (FSRQs) and BL~Lac objects are 
active galactic nuclei (AGNs) commonly unified in the class 
of blazars. They exhibit some of the most violent high-energy
phenomena observed in AGNs to date. Their spectral energy
distributions (SEDs) are characterized by non-thermal continuum 
spectra with a broad low-frequency component in the radio -- UV 
or X-ray frequency range and a high-frequency component from
X-rays to $\gamma$-rays. 
In the framework of relativistic jet models, the low-frequency (radio
-- optical/UV) emission from blazars is interpreted as synchrotron
emission from nonthermal electrons in a relativistic jet. The
high-frequency (X-ray -- $\gamma$-ray) emission could either be
produced via Compton upscattering of low frequency radiation by the
same electrons responsible for the synchrotron emission \citep[leptonic
jet models; for a recent review see, e.g.,][]{boettcher07a}, or 
due to hadronic processes initiated by relativistic protons 
co-accelerated with the electrons \citep[hadronic models, for 
a recent discussion see, e.g.,][]{muecke01,muecke03}. 

The quasar 3C279 ($z = 0.536$) is one of the best-observed flat 
spectrum radio quasars, in part because of its prominent 
$\gamma$-ray flare shortly after the launch of the {\it Compton
Gamma-Ray Observatory (CGRO)} in 1991. It was persistently 
detected by the {\it Energetic Gamma-ray Experiment Telescope 
(EGRET)} on board {\it CGRO} each time it was observed, even in 
its very low quiescent states, e.g., in the winter of 1992 -- 1993, 
and is known to vary in $\gamma$-ray flux by roughly two orders 
of magnitude \citep{maraschi94,wehrle98}. It has been monitored
intensively at radio, optical, and more recently also X-ray
frequencies, and has been the subject of intensive multiwavelength
campaigns \citep[e.g.,][]{maraschi94,hartman96,wehrle98}. 
Observations with the {\it International Ultraviolet Explorer} 
in the very low activity state of the source in December 
1992 -- January 1993 revealed the existence of a thermal 
emission component, possibly related to an accretion disk
\citep{pian99}. 

A complete compilation and modeling of all available SEDs simultaneous 
with the 11 {\it EGRET} observing epochs has been presented in \cite{hartman01}. 
The modeling was done using the time-dependent leptonic synchrotron
self-Compton (SSC) + External Compton (EC) model of \cite{bms97,bb00} 
and yielded quite satisfactory fits for all epochs. The results were 
consistent with other model fitting works 
\citep[e.g.,][]{bednarek98,sikora01,moderski03} concluding that the 
X-ray -- soft $\gamma$-ray portion of the SED might be dominated by 
SSC emission, while the {\it EGRET} emission might require an additional
component, most likely external-Compton emission. 

During a recent observing campaign by the Whole Earth Blazar Telescope
(WEBT) collaboration \citep{boettcher07b} in the spring of 2006, intensive 
monitoring by the Major Atmospheric Gamma-Ray Imaging Cherenkov Telescope 
(MAGIC) yielded a positive detection at $> 100$~GeV on February 23, 2006
\citep{albert08}. This makes 3C279 the first quasar and (as of April 2009) 
the most distant object detected in VHE $\gamma$-rays. In this paper, we 
present the optical (BVRI) and X-ray ({\it RXTE}) data taken simultaneously 
with the MAGIC detection, and discuss the implications of this detection 
for current standard blazar jet models. 

Throughout this paper, we refer to $\alpha$ as the energy 
spectral index, $F_{\nu}$~[Jy]~$\propto \nu^{-\alpha}$. A 
cosmology with $\Omega_m = 0.3$, $\Omega_{\Lambda} = 0.7$, 
and $H_0 = 70$~km~s$^{-1}$~Mpc$^{-1}$ is used. In this cosmology
the luminosity distance of 3C~279 at a redshift of $z = 0.536$ 
is $d_L = 3.08$~Gpc. 

\begin{figure}
\plotone{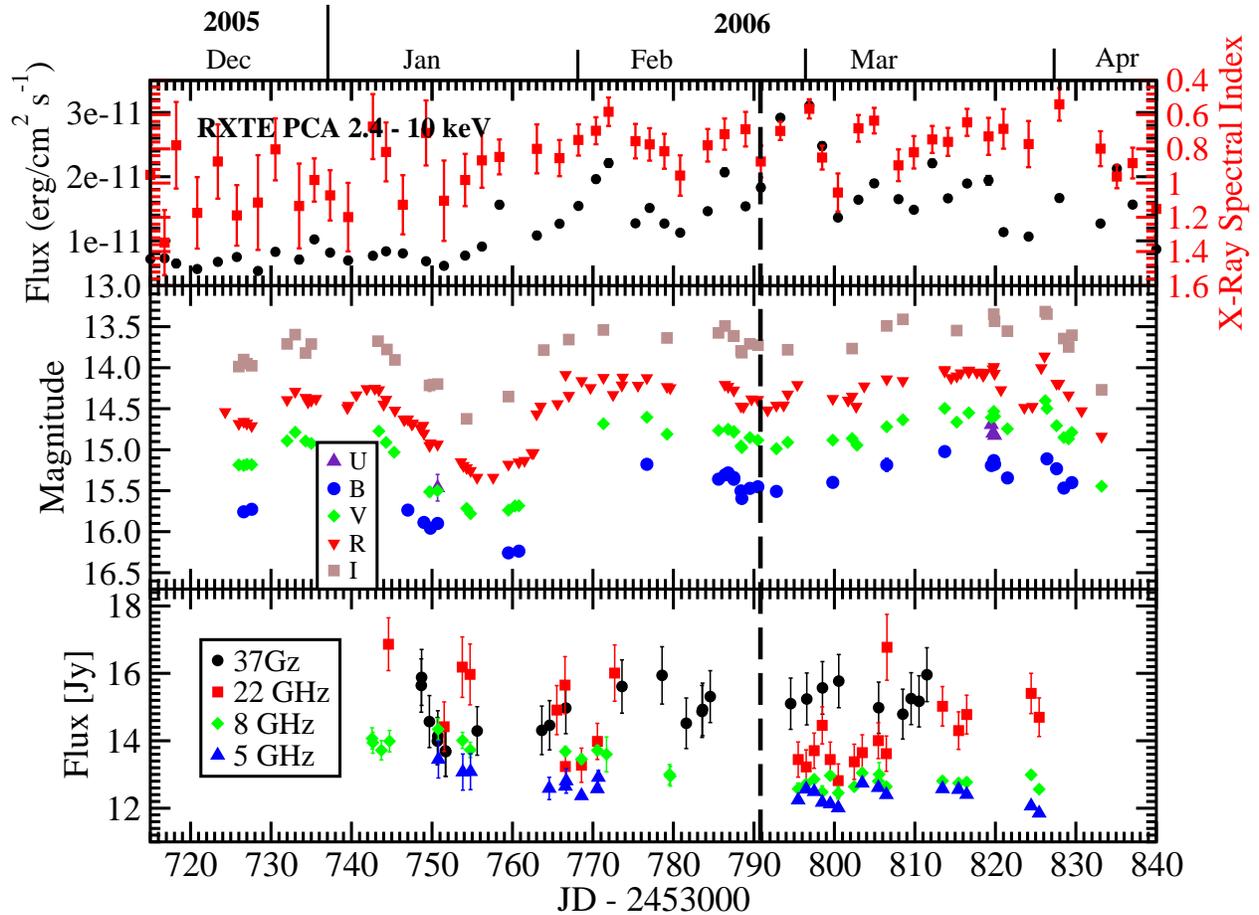}
\caption{Light curves at radio (bottom), optical (center), and X-ray
frequencies over the course of the WEBT campaign in the spring of 2006.
The red points (right axis labels) indicate the energy spectral 
index in the 2 -- 10~keV range. The vertical dashed line marks the day 
of the MAGIC VHE $\gamma$-ray detection.}
\label{lightcurves}
\end{figure}

\section{\label{observations}Observations and results}

3C~279 was observed in a WEBT campaign at radio, near-IR, and optical 
frequencies, throughout the spring of 2006. Details of the observations,
data analysis, and implications of the optical variability patterns
observed during that campaign have been published in \cite{boettcher07b}.
The source was simultaneously monitored with 3 pointings per week with
the {\it Rossi X-ray Timing Explorer (RXTE)} Proportional Counter Array 
(PCA). We obtained the X-ray flux measurements with the PCA detector PCU2, 
using typical exposure times of 2~ks for each pointing. The data reduction
is described in \cite{chat08}.

Fig. \ref{lightcurves} shows the radio, optical and X-ray light curves 
of 3C279 during spring 2006, along with the 2 -- 10~keV energy
spectral index as a function of time. The dashed vertical line marks 
the day of the MAGIC $> 100$~GeV $\gamma$-ray detection. While the 
source was overall in an extended optical high state ($R \sim 14.5$), 
no extraordinary variability in any optical (BVRI) band was observed at the 
time of the MAGIC detection. 

During most of December 2005 and January 2006, the X-ray flux of 3C279
was in a low state, near its historical minimum. Around Jan. 25, however,
the source made a transition to a higher X-ray flux state with substantial
variability in flux and spectral index on a characteristic time scale of
$\sim 10$~days. The average flux increased to about a factor $\sim 2$ 
-- 3 compared to the low state. In the high state, there is a clear 
correlation between X-ray flux and spectral hardness, with the spectrum 
becoming harder as the flux increases. Statistical uncertainties preclude 
any conclusions about a flux-hardness correlation in the low X-ray state. 
The VHE flare observed by MAGIC precedes an X-ray outburst with the 
highest X-ray flux measured since the major optical/X-ray outburst in
2001 \citep[see, e.g.][]{chat08}, by $\sim 5$ -- 7~days. 

\begin{figure}
\plotone{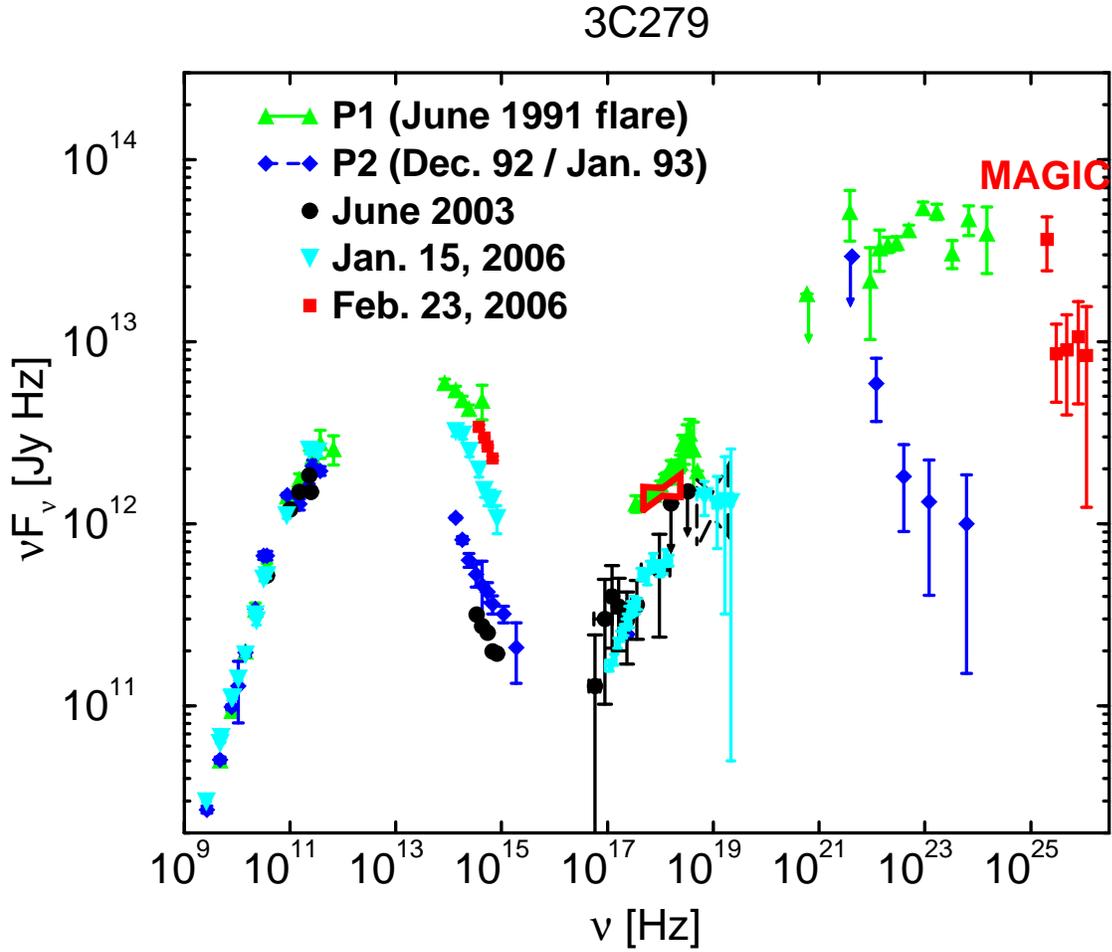}
\caption{Compilation of broadband spectral energy distributions of
3C279, including the day of the MAGIC VHE detection, February 23, 2006.
Data for P1 and P2 are from \cite{hartman01}, 2003 data are from
\cite{collmar04}, and 2006 data are from \cite{collmar07} and 
\cite{boettcher07b}. }
\label{SED}
\end{figure}

Fig. \ref{SED} compares several historical SEDs of 3C279 to the one
measured on February 23, 2006, along with the MAGIC VHE detection. 
The X-ray flux is comparable to the one observed during the major
EGRET-detected $\gamma$-ray outburst in June 1991. The 
MAGIC data points show the measured flux, corrected for intergalactic
$\gamma\gamma$ absorption, using the lowest plausible level of 
extragalactic background light (EBL) cosmic infrared light, 
according to the model of \cite{primack05}. As discussed in 
\cite{albert08}, this yields a best-fit energy index to the
corrected VHE spectrum of $\alpha_{\rm VHE} = 1.9 \pm 0.9_{\rm stat}
\pm 0.5_{\rm syst}$. Other currently discussed EBL models would 
predict a higher intrinsic VHE $\gamma$-ray flux and substantially
harder spectrum. In particular, a high-EBL model \citep{stecker06}
would lead to an intrinsic VHE spectral index of $\alpha_{\rm VHE} = 
-0.5 \pm 1.2_{\rm stat} \pm 0.5_{\rm syst}$. As we will discuss below,
the VHE spectrum corrected with the \cite{primack05} model already
poses severe constraints on currently discussed blazar models. 
Those constraints would be even more restrictive and problematic
for a higher EBL model. In particular, the high-EBL spectral index
quoted above would locate the high-energy peak of the SED of 3C~279
at TeV energies, which would lead to even more unrealistic parameter
choices for the leptonic models considered in the following section. 
We will therefore restrict our more detailed discussion to the VHE 
spectrum corrected by the low EBL model.
The optical spectrum, while clearly in an elevated state, shows
about the same, steep spectral index $\alpha_{\rm opt} \sim 1.7$ 
as during lower optical flux states, indicating an underlying
nonthermal electron spectral index of $p = 4.4$. 

The slopes of the radio and optical spectra indicate that the 
synchrotron peak was in the usual range where it has been observed
in many previous observing campaigns, i.e., in the infrared regime,
around $\nu_{\rm sy} \sim 5 \times 10^{13}$~Hz, corresponding to a 
dimensionless photon energy $\epsilon_{\rm sy} \equiv h \nu_{\rm sy} 
/ (m_e c^2) \sim 4 \times 10^{-7}$. This is consistent with the location 
of the synchrotron peak in a compilation of simultaneous multiwavelength 
data in mid-2006, shown by \cite{marscher08}, which included infrared 
coverage by the {\it Spitzer Space Telescope}. For the purpose of a 
quantitative analysis, the synchrotron peak flux may be estimated to 
be of the order of $\nu F_{\nu}^{\rm sy} \sim 10^{13}$~Jy~Hz. Equally,
the X-ray spectrum shows a quite typical shape as observed in previous
high states of 3C~279, in particular the P1 SED shown in Figs. \ref{SED}
and \ref{fit}. This suggests that the X-ray -- GeV $\gamma$-ray
spectrum is similar to previously observed high states during the EGRET
era.

If a one-zone leptonic jet model (as discussed in the following section) 
applies, the VHE spectrum is expected to be at least as steep as 
the optical (synchrotron) flux. The spectral indices are expected to
be similar if the $\gamma$-ray emission is produced by Compton scattering
in the Thomson regime. If Klein-Nishina effects are important in the
production of VHE $\gamma$-rays, the resulting VHE spectrum would be
even steeper than the synchrotron spectrum. As already indicated above,
this would be in direct conflict with the observed relatively hard
intrinsic VHE spectrum, even when corrected with a low EBL model.
Therefore, in order not to predict a 
GeV $\gamma$-ray flux greatly in excess of any archival EGRET flux, it 
is reasonable to assume a $\gamma$-ray peak 
at $\nu_{\gamma} \sim 10^{24}$ -- $10^{25}$~Hz,
corresponding to $\epsilon_{\gamma} \sim 10^5$. The $\gamma$-ray 
peak flux is then $\nu F_{\nu}^{\gamma} \sim 5 \times 10^{13}$~Jy~Hz. 
Previous modeling works of the SEDs of 3C~279 placed the peak
of the high-energy component typically at frequencies around
$\nu_{\gamma} \sim 10^{23}$~Hz, i.e., about 1 -- 2 orders of magnitude 
lower than our new estimate, taking the MAGIC results into account.
It is this shift of the inferred high-energy peak which will lead to 
quite dramatically different model implications for both leptonic and 
hadronic models compared to previous modeling efforts.

From the spectral upturn in the UV in the P2-spectrum in 
Fig. \ref{SED}, we can estimate a thermal (external) photon
source with a luminosity of $L_D \sim 2 \times 10^{45}$~erg~s$^{-1}$,
peaking at $\nu_D \sim 10^{15}$~Hz ($\epsilon_D \sim 10^{-5}$).

\section{\label{leptonic}Implications for Leptonic Jet Models}

In this section, we consider whether a one-zone leptonic blazar 
jet model can account for the February 23, 2006, SED of 3C279.
We consider a scenario in which a nonthermal population of ultrarelativistic
electrons produces, at the same time, the synchrotron emission from
radio through UV and the $\gamma$-ray emission via Compton scattering
of soft photons off the relativistic electrons. In general, we will
assume that electrons are accelerated into a power-law distribution
in electron energy, $Q(\gamma) = Q_0 \, \gamma^{-q}$, in the range
$\gamma_1 \le \gamma \le \gamma_2$. The emission region has a radius 
$R_B \equiv 10^{16} \, R_{16}$~cm. We define an escape time scale 
parameter $\eta_{\rm esc}$ such that the escape time scale for 
relativistic electrons is $\tau_{\rm esc} \equiv \eta_{\rm esc} 
\, R_B/c$. The interplay between radiative
cooling and escape leads to the development of a spectral break in
the electron spectrum at a Lorentz factor $\gamma_b$, where the 
radiative cooling time scale equals the escape time scale. If the 
primarily injected electron distribution has a low-energy cutoff 
below the break Lorentz factor $\gamma_b$ (the slow cooling regime), 
the spectral index of the electron distribution at $\gamma < \gamma_b$ 
is $p = q$, while above it the spectrum is steepened to $p = q + 1$. 
However, given the steep spectral index of the optical spectrum, 
implying $p = 4.4$ in the electron energy range synchrotron-radiating 
in the optical regime, a spectral break from $p = 3.4$ to $p = 4.4$ 
would not produce a peak in the $\nu F_{\nu}$ spectrum at the 
characteristic synchrotron frequency corresponding to $\gamma_b$. 
Therefore, it is more likely that the injected electron distribution 
has a high low-energy cutoff at $\gamma_1 > \gamma_b$. In this case, 
particles at energies $\gamma < \gamma_1$ result only from radiative
cooling from higher energies, resulting in an electron spectrum with
a spectral index $p = 2$ in the range $\gamma_b \le \gamma \le \gamma_1$, 
while above $\gamma_1$, we have $p = q + 1$, as in the slow cooling
case. 

\subsection{\label{SSC}SSC model}

In the past, VHE emission has only been observed in high-frequency 
peaked BL Lac objects. In that case, a model interpreting the 
$\gamma$-ray emission as synchrotron self-Compton (SSC) emission 
has proven to be very successful, although recent observations of 
rapid variability on time scales of a few minutes (\cite{aharonian07} 
for PKS 2155-304, and \cite{albert07} for Mrk 501) are posing serious 
challenges to this interpretation \citep[e.g.][]{finke08}, and the 
VHE emission of the intermediate BL~Lac object W Comae, recently 
detected by VERITAS \citep{beilicke08} is more plausibly explained 
by Comptonization of an external radiation field. 
We point out that previous modeling efforts on 3C279, prior 
to the MAGIC detection, have concluded that an EC component is 
strongly preferred to explain the GeV $\gamma$-ray detection 
\citep{hartman01,bednarek98,sikora01,moderski03}. However, since the MAGIC
points provide a yet unexplored new constraint on blazar models for
3C279, it is worthwhile to revisit the SSC hypothesis in this paper.

Given the synchrotron origin of the low-frequency peak at $\epsilon_{\rm sy} 
\sim 4 \times 10^{-7}$ and the SSC origin of the $\gamma$-ray peak
at $\epsilon_{\gamma} \sim 10^5$, the Lorentz factor of electrons $\gamma_p$ 
radiating at the synchrotron and SSC peaks, can be estimated from

\begin{equation}
\gamma_p = \sqrt{\epsilon_{\gamma} \over \epsilon_{\rm sy}} \sim
1.6 \times 10^5.
\label{gp}
\end{equation}

At the same time, the synchrotron peak frequency is given by

\begin{equation}
\nu_{sy} = 4.2 \times 10^6 \, \gamma_p^2 \, B_G \, D / (1 + z) \; {\rm Hz}
\label{nusy}
\end{equation}
where $B_G$ is the (co-moving) magnetic field in Gauss, and
$D \equiv 10 \, D_1 = (\Gamma [ 1 - \beta_{\Gamma} 
\cos\theta_{\rm obs}])^{-1}$ is the Doppler enhancement 
factor. This yields an estimate of the magnetic field and 
the Doppler factor of

\begin{equation}
B_G \, D_1 \sim 7 \times 10^{-5}.
\label{BD}
\end{equation}
This indicates that such a scenario would imply unrealistically
low magnetic fields compared to standard values of $\sim 1$~G found
from SED modeling of 3C279 in other states, as well as other blazar-type
quasars. 
Even other known TeV blazars, whose SEDs can usually be well
represented with SSC models, typically require magnetic fields of
$B \gtrsim 0.1$~G, several orders of magnitude above the estimate
found here for 3C279. We therefore conclude that a one-zone SSC model 
for the SED of 3C279 on February 23, 2006, including the VHE $\gamma$-ray 
emission is very problematic.

\subsection{\label{EC}External Compton}

The leptonic external-Compton scenario is based on the assumption that 
photons from an external, quasi-isotropic radiation field with dimensionless
photon energy $\epsilon_s$ are Compton-upscattered to the observed 
$\gamma$-ray photon energies. As mentioned earlier, if Klein-Nishina 
effects become important in the production of the VHE emission, the 
resulting VHE $\gamma$-ray spectrum would be even steeper than the 
observed synchrotron spectrum, in contradiction with the observed 
VHE spectrum, even when corrected for the lowest plausible level 
of EBL $\gamma\gamma$ absorption. We therefore consider here only 
the possibility that the VHE $\gamma$-rays in a leptonic scenario 
are produced by Thomson scattering.
Soft photons can be upscattered effectively in the Thomson regime 
at most up to energies $\epsilon_{\gamma} = 1 / \epsilon_s$. This
indicates that a photon field with a characteristic
photon energy of the accretion disk field at $\epsilon_D \sim 10^{-5}$
can effectively serve as the seed photon field for upscattering to
the observed $> 100$~GeV $\gamma$-rays. We assume that a fraction
$\tau_{BLR} \equiv 10^{-1} \tau_{-1}$ of the accretion disk radiation
is reprocessed in the broad line region, which is located at an average
distance $R_{\rm BLR} \equiv 0.1 \, R_{\rm BLR, -1}$~pc from the central 
engine. HST near-UV spectroscopy \citep{pian05} indicates that the total 
luminosity of the BLR in 3C279 is $L_{\rm BLR} \sim \tau_{\rm BLR} \, L_D 
\sim 2 \times 10^{44}$~erg/s, motivating the above scaling in terms of
$\tau_{-1}$. In the co-moving frame, the external photons will thus have 
a characteristic energy of ${\epsilon'}_s = \Gamma \, \epsilon_s$. In 
addition to Eq. (\ref{nusy}), we now have an independent estimate for 
$\gamma_p$, namely

\begin{equation}
\gamma_p = \sqrt{\epsilon_{\gamma} \over \Gamma^2 \epsilon_D} \sim 10^4 
\Gamma_1^{-1}.
\label{gpec}
\end{equation}

We can use Eq. (\ref{nusy}) to obtain an estimate for the magnetic
field:

\begin{equation}
B_G = 1.8 \times 10^{-2} \, \Gamma_1^2 \, D_1^{-1}.
\label{Bsy}
\end{equation}

The energy density of external photons in the co-moving frame can be
expressed as

\begin{equation}
{u'}_{\rm ext} \sim {L_D \, \tau_{\rm BLR} \, \Gamma^2 \over 
4 \pi R_{\rm BLR}^2 
\, c}.
\label{uext}
\end{equation}

From the $\gamma$-ray dominance, $L_{\gamma} / L_{\rm sy} \sim {u'}_{\rm ext}
/ {u'}_B \sim 5$, we can then obtain an independent estimate of the magnetic
field of

\begin{equation}
B_G = 1.0 \, \tau_{-1}^{1/2} \, \Gamma_1 \, R_{\rm BLR, -1}^{-1}.
\label{Bgamma}
\end{equation}
Combining the magnetic field estimates (\ref{Bsy}) and (\ref{Bgamma}),
we find

\begin{equation}
R_{\rm BLR, -1} = 57 \, \tau_{-1}^{1/2}
\label{RBLR_B}
\end{equation}
which is in drastic contrast to the estimate of \cite{pian05} of
$R_{\rm BLR} \sim 3 \times 10^{-2}$~pc.

Considering the peak level of the synchrotron flux, we can use
Eq. (8) of \cite{boettcher03} to relate the magnetic field in the
emission region with the equipartition fraction $e_B \equiv {u'}_B
/ {u'_e}$, i.e., the ratio of co-moving energy densities in the 
magnetic field and the nonrelativistic electron population:

\begin{equation}
B_{e_B} = 1.25 \, D_1^{-1} \left({d_{27}^4 \, f_{-10}^2 \, e_B^2 \over
[1 + z]^4 \, \epsilon_{\rm sy, -6} \, R_{16}^6 \, [p - 2]} \right)^{1/7}.
\label{BeB}
\end{equation}
Setting this equal to the magnetic-field estimate (\ref{Bsy}) yields

\begin{equation}
e_B = 9.7 \times 10^{-9} \, R_{16}^3 \, \Gamma_1^7.
\label{eB}
\end{equation}
Consequently, if we choose a conventional value of the Lorentz factor
$\Gamma \sim 15$, we find an uncomfortably low magnetic field of $B \sim
0.03$~G, corresponding to $e_B \sim 1.7 \times 10^{-7} \, R_{16}^3$, i.e., 
a far sub-equipartition magnetic field. Such a situation would make jet
confinement very problematic, and is in contradiction with model results
for 3C279 in other observing epochs and for other quasar-type blazars in
general, where magnetic fields of typically $B \sim 1$ -- a few G are 
inferred, in approximate equipartition with the relativistic electron
population.

Alternatively, forcing the system to attain approximate equipartition
would require us to assume an uncomfortably high Lorentz factor of
$\Gamma \sim 140 \, R_{16}^{-3/7}$. This choice of a bulk Lorentz 
(and Doppler) factor would imply $B \sim 0.25$~G, and a low-energy 
cut-off of the injected electron population at $\gamma_1 = \gamma_p 
\sim 710 \, R_{16}^{3/7}$. Apart from the fact that this is an 
order of magnitude larger than bulk Lorentz factors inferred from 
superluminal motion, it would require an implausibly close alignment 
of the jet with our line of sight, $\theta_{\rm obs} \sim 0.4^o$. 
We note that the magnetic-field estimate of Eq. (\ref{Bsy}) carries 
a proportionality $B \propto (\epsilon_s / \epsilon_{\gamma})$. 
Since our choice of $\epsilon_s \sim 10^{-5}$ is already close 
to the largest possible value to allow Thomson scattering to 
TeV $\gamma$-rays, the assumption of a different soft photon 
source (necessarily with a smaller $\epsilon_s$) would worsen 
the problem of the unusually small inferred magnetic field.

\begin{figure}
\plotone{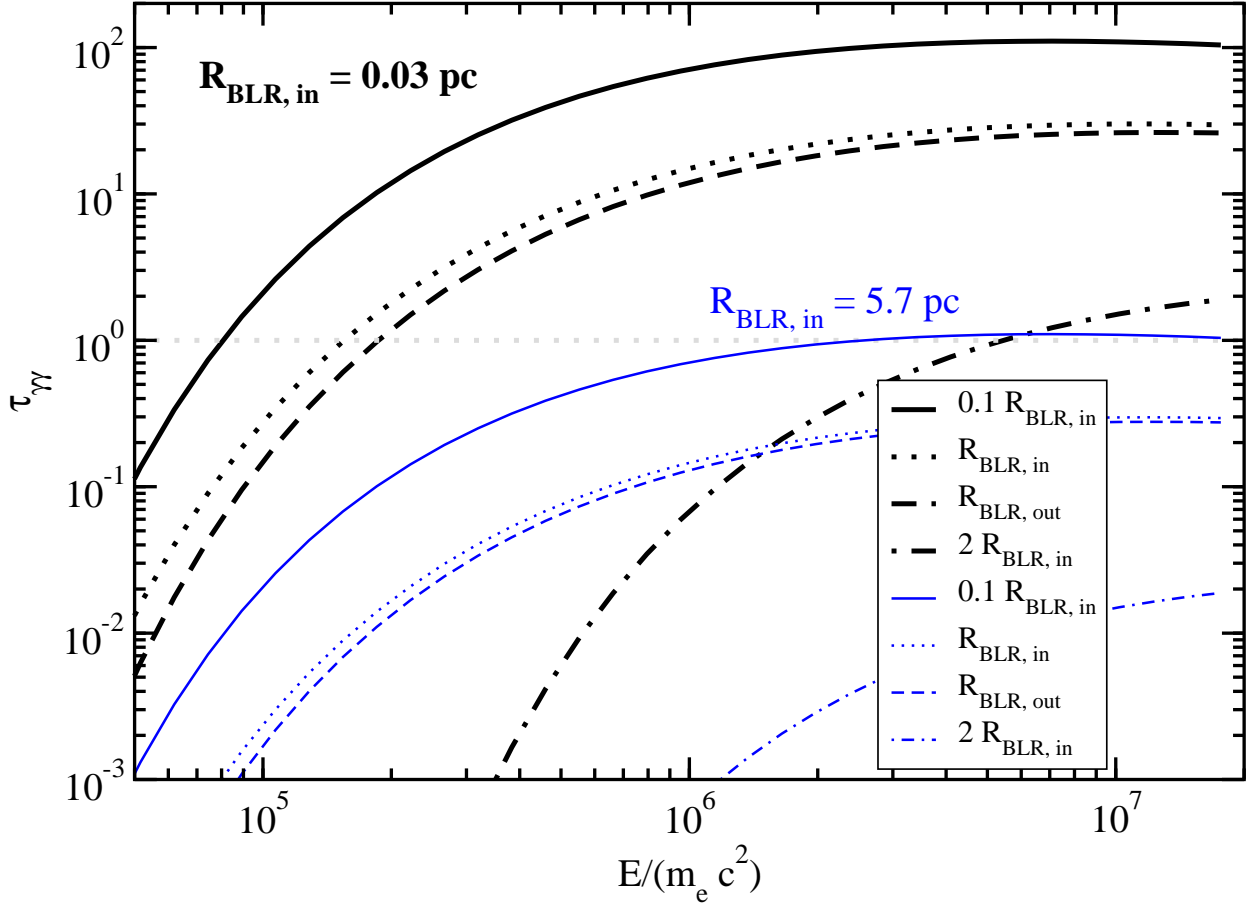}
\caption{Opacity for VHE $\gamma$-ray photons due to $\gamma\gamma$ absorption
on the BLR radiation field. The labels denote the location of the $\gamma$-ray
emitting region. Other parameters: $L_D = 2 \times 10^{45}$~erg/s, $\Theta_D = 
10^{-5}$, $\tau_{\rm BLR} = 0.1$. Heavy (black) curves refer to the value of
$R_{\rm BLR, in} = 0.03$~pc as inferred by \cite{pian05} with the outer edge
of the BLR, $R_{\rm BLR, out} = 0.031$~pc; light (blue) curves refer to 
$R_{\rm BLR, in} = 5.7$~pc as inferred from Eq. \ref{RBLR_B} and $R_{\rm BLR, out} 
= 5.8$~pc. The photon energy $E$ is in the stationary AGN rest frame. }
\label{taugg}
\end{figure}

It has been noted by several authors that the $\gamma\gamma$ absorption
of VHE $\gamma$-ray photons by the radiation field of the BLR may present
another problem for a model of VHE $\gamma$-ray emission inside the BLR of 
luminous quasars in general \citep[e.g.][]{dp03,reimer07} and 3C279 in particular 
\citep{liu08,sb08}. We therefore need to investigate the effects of $\gamma\gamma$ 
absorption in the BLR radiation field for the parameters we inferred 
above. In \cite{bd95}, the time-dependent $\gamma\gamma$ absorption 
signatures of an accretion disk flare (reflected by BLR clouds) on 
VHE $\gamma$-ray emission were investigated. We have modified their 
approach for our purposes, adopting non-variable accretion-disk 
emission. The standard optically-thick accretion-disk spectrum is
approximated by a spectral shape $F_{\epsilon} \propto 
\epsilon^{1/3} \, e^{-\epsilon / \Theta_D}$, where $\Theta_D \sim 10^{-5}$ 
is the dimensionless inner disk temperature, $\Theta_D = kT_{\rm D, in} / 
(m_e c^2)$. Fig. \ref{taugg} illustrates the dependence of the $\gamma\gamma$ 
absorption depth as a function of the dimensionless photon energy $\epsilon$ 
and location of the $\gamma$-ray production site. Our results confirm the
findings of \cite{liu08}: With standard values of the BLR parameters , 
$\tau_{\rm BLR} \sim 0.1$, $R_{\rm BLR} \sim 0.03$~pc, as inferred by
\cite{pian05}, VHE $\gamma$-rays produced within the BLR of 3C279 suffer 
severe $\gamma\gamma$ absorption by the same photon field that would 
serve as seed photon field for Compton scattering in a leptonic model. 
For the extreme parameters of $\Gamma \sim 140$, requiring $R_{\rm BLR} 
\sim 5.7$~pc, $\tau_{\rm BLR} \sim 0.1$, $\gamma\gamma$ absorption would
hardly be a problem even out to multi-TeV $\gamma$-ray energies, if the
$\gamma$-rays are produced close to the inner boundary of the BLR.
Any level of $\gamma\gamma$ absorption in the BLR radiation field 
would imply an even higher level of intrinsic VHE $\gamma$-ray emission
and therefore make the requirements for jet parameters in a one-zone leptonic
model even more extreme. 

\begin{figure}
\plotone{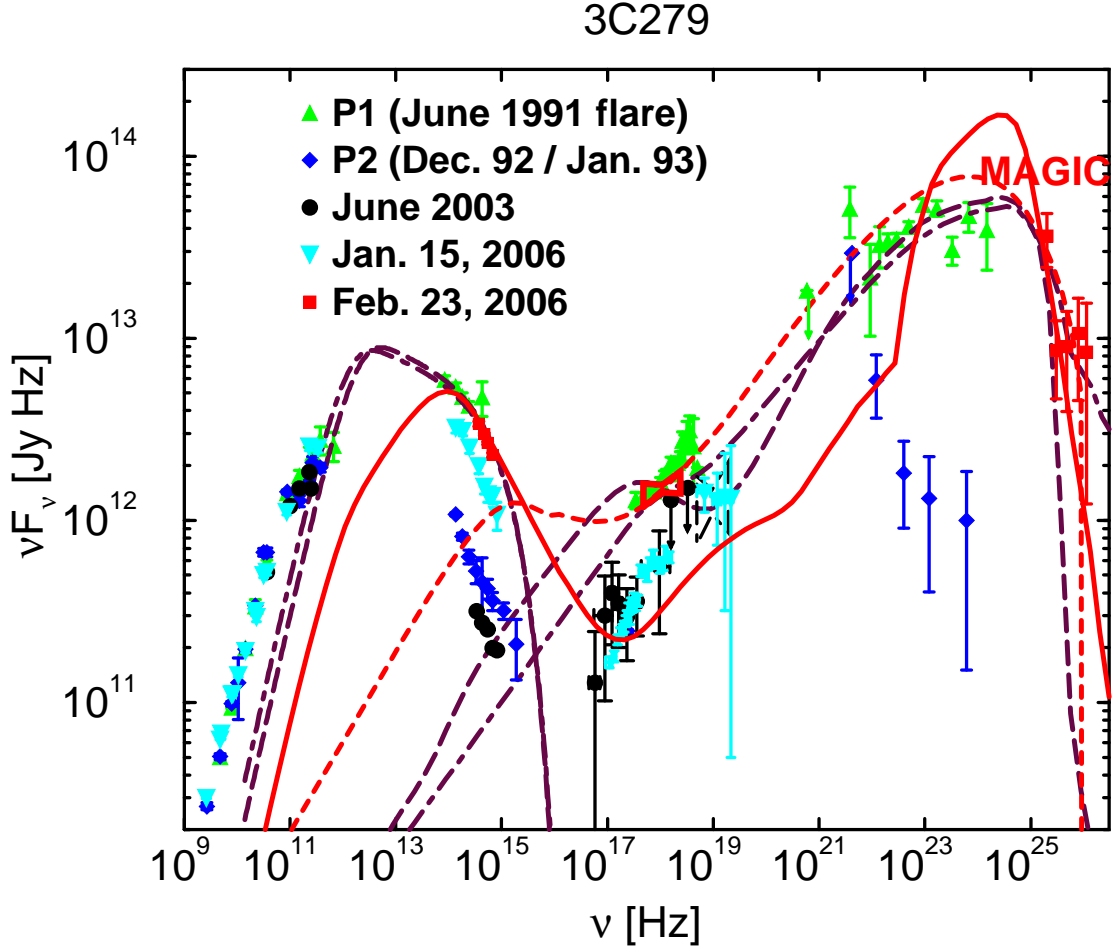}
\caption{Spectral fits to the SED of 3C279 on February 23, 2006:
(solid [red]) using a leptonic External-Compton model with parameters 
similar to those derived in \S \ref{EC}; (short-dashed [red]) leptonic 
SSC model fit only to the X-ray -- $\gamma$-ray spectrum. Relevant 
parameters: $\gamma_1 = 10^4$, $\gamma_2 = 10^6$, $q = 2.3$, $L_{\rm j, e} 
= 2.2 \times 10^{47}$~erg~s$^{-1}$, $\eta_{\rm esc} = 80$, $R_B =
6 \times 10^{15}$~cm, $\Gamma = D = 20$, $B = 0.2$~G; (dot-dashed 
[maroon]) fit with the hadronic synchrotron-proton blazar model 
with internal (synchyrotron) photons only as targets for p$\gamma$ 
pion production, and (long-dashed [maroon]) with synchrotron +
external (BLR) photons as targets for p$\gamma$ pion production.
See table \ref{SPBparameters} for parameters.}
\label{fit}
\end{figure}

The red, solid curve in Fig. \ref{fit} shows a leptonic model 
calculation with parameters similar to the quasi-equipartition 
case outlined above. We used an equilibrium version of the 
time-dependent SSC + EC model of \cite{bc02}. While the 
optical (synchrotron) spectrum and the level of the VHE $\gamma$-ray 
flux can be reproduced reasonably well, it is obvious that the X-ray 
flux is grossly underproduced. This is a consequence of the required,
rather large low-energy cutoff at $\gamma_1 \sim 700$. The choice
of a substantially lower $\gamma_1$ would extend both the synchrotron 
and the $\gamma$-ray spectra towards lower frequencies along the slopes
of the optical and VHE $\gamma$-ray spectra and would therefore produce 
unreasonably large infrared and MeV -- GeV $\gamma$-ray fluxes. In
addition, it would require a larger jet power and therefore drive
the system further out of equipartition (towards far sub-equipartition
magnetic fields). Also, even though most of the $> 100$~GeV flux is produced by
Compton scattering in the Thomson regime, the resulting VHE $\gamma$-ray
spectrum appears to be steeper than the observed one, even with 
correction for the low EBL level according to \cite{primack05}.

We therefore conclude that both the SSC and the external-Compton
scenario for a one-zone, homogeneous jet model face severe problems 
representing the simultaneous SED of 3C279 on February 23, 2006, 
including the VHE $\gamma$-ray emission.

As noted earlier, the VHE $\gamma$-ray flare was not accompanied
by any remarkable optical variability. This may be another hint 
that in 3C279 the optical and VHE $\gamma$-ray fluxes may be produced 
in separate emission regions. However, the calculation of the
$\gamma\gamma$ opacity above indicates that even an inhomogeneous
leptonic jet model would face severe problems in an external-Compton
scenario. The dashed red curve in Fig. \ref{fit} indicates that
an SSC model can successfully reproduce the X-ray -- VHE $\gamma$-ray
spectrum with reasonable parameters, but fails to reproduce the
optical spectrum. This may indicate support for a multi-zone leptonic
model. This would require that the VHE $\gamma$-ray emission is 
produced far outside the BLR, possibly in an internal shock scenario
\citep{spada01,sokolov04}.
This is indicated both by the $\gamma\gamma$ opacity argument as well
as the low required magnetic field for the SSC fit presented in Fig.
\ref{fit}. Such a model would imply that the X-ray emission is produced
by low-energy electrons which have undergone substantial radiative
cooling. One would therefore predict a delay of X-rays with respect
to $\gamma$-rays on the order of the electron cooling time scale,

\begin{equation}
\tau_{\rm cool, SSC}^{\rm obs} \sim {m_e c^2 \over D \, (4/3) \, c \, 
\sigma_T \, {u'}_B \, (L_{\rm SSC}/L_{\rm sy}) \, \gamma}.
\label{taussc}
\end{equation}
Using the value of $B = 0.2$~G used for the model shown in Fig. \ref{fit},
and scaling the electron energy $\gamma \equiv 10^3 \, \gamma_3$, we find
the expected $\gamma$-ray vs. X-ray delay as $\tau \sim 27 \, \gamma_3^{-1}$~hr.
We note, however, that an intensive search for inter-band time delays between
optical, X-ray and $\gamma$-ray emission of 3C279 during the EGRET era by 
\cite{hartman01b} did not find any evidence for systematic $\gamma$-ray vs. 
X-ray delays.

\section{\label{hadronic}Hadronic Models}

If relativistic protons at energies above the threshold for p$\gamma$
pion production are present in the jet of 3C~279, hadronic interactions 
must be considered in blazar emission models. For the present modeling 
we use the hadronic Synchrotron-Proton Blazar (SPB) model of 
\cite{muecke01,muecke03}. In its original version, this model was
best suited for X-ray selected BL~Lac objects with very weak or
absent external radiation fields. However, the quasar 3C~279 is known 
to have strong accretion disk and BLR line emission (see \S \ref{EC}).
We therefore extend the (one-zone) SPB model to account also for target 
photon fields external to the jet.

Relativistic electrons (e) and protons (p) with a power law index 
$q_e = q_p$ are injected instantaneously into a spherical 
emission region, or blob, which is moving with relativistic 
velocity along the jet axis. In the strongly magnetized (with field 
strength $B =$~const.) blob, the primary electrons lose their energy 
predominantly through the synchrotron channel. The resulting synchrotron 
radiation generally dominates the low energy component of the blazar SED, 
and serves as target photon field for proton-photon interactions and 
pair (synchrotron) cascading. In our extension of the SPB model we add 
the photon field from the BLR, simplified as an isotropic distribution 
in the jet frame \citep[see][for a discussion of this approximation]{reimer05} 
as an additional target for particle-photon interactions and cascading. 
We properly take into account the Doppler boost of the average
photon energy and photon energy density of the external photon field. 
In the present work we use the approximation
advocated by \cite{tg08} for the BLR radiation field: In the co-moving 
jet frame a blackbody spectrum with temperature $T' \propto \Gamma 
\nu_{Ly\alpha}$ ($\nu_{Ly\alpha}$ the Ly$\alpha$ line frequency) is 
used to approximate the boosted BLR emission from the stationary AGN
to the blob frame. Local absorption of $\gamma$-rays \citep{reimer07} 
external to the emission region is taken into account as well.
The injected relativistic protons suffer energy losses from photomeson 
production, Bethe-Heitler pair production, synchrotron radiation and 
adiabatic expansion. Charged particles produced as secondaries in the 
photomeson production channel may suffer synchrotron losses as well 
prior to their decay. This is particularly relevant for charged 
pions and muons (henceforth named $\mu$-synchrotron radiation). All
high energy photons may initiate pair cascades, which redistributes a
fraction of the photon power from high to lower energies where the 
photons can eventually escape the emission region.

\begin{deluxetable}{cccc}
\tabletypesize{\scriptsize}
\tablecaption{Parameters for the hadronic synchrotron-proton blazar
model fits to 3C279 (Figs. \ref{fit} and \ref{hfit}) for (2nd col.) internal synchrotron
photons only, and (3rd col.) internal synchrotron + external (BLR) photons
as targets for p$\gamma$ pion production. }
\tablewidth{0pt}
\tablehead{
\colhead{Parameter} & \colhead{symbol} & \colhead{SPB (sy.)} & \colhead{SPB (sy. + ext.)} }
\startdata
Jet power 	& $L_j$ [erg $s^{-1}$]	& $1 \times 10^{49}$ 		& 6.6 $\times 10^{48}$ \\
Doppler factor	& $D$			& 12				& 12 \\
Blob radius	& $R_b$			& $3 \times 10^{16}$~cm 	& $3 \times 10^{16}$~cm \\
Maximum proton energy & $E_{\rm p, max}$ & 5 $\times 10^9$~GeV		& $4.5 \times 10^9$~GeV \\
Magnetic field	& $B$			& 40~G				& 60~G \\
Particle spectral index & $q_p = q_e$	& 2.2				& 2.2 \\
Proton energy density & $u_p$ [erg cm$^{-3}$]	& $1.2 \times 10^3$	& 540 \\
Electron/proton density ratio & $n_e/n_p$	& $2 \times 10^{-3}$	& $3 \times 10^{-3}$\\
Temperature of BLR emission & $kT_{\rm BLR} $	& ---			& 3 eV \\
BLR photon energy density & $u_{\rm BLR}$	& ---			& $10^{-3}$~ergs~cm$^{-3}$ \\
BLR radius 	& $R_{\rm in, BLR}$		& ---			& 0.1~pc \\
Location of emission region & $R_{\rm em}$ 	& ---			& 0.18~pc \\
\enddata
\label{SPBparameters}
\end{deluxetable}

\begin{figure}
\plottwo{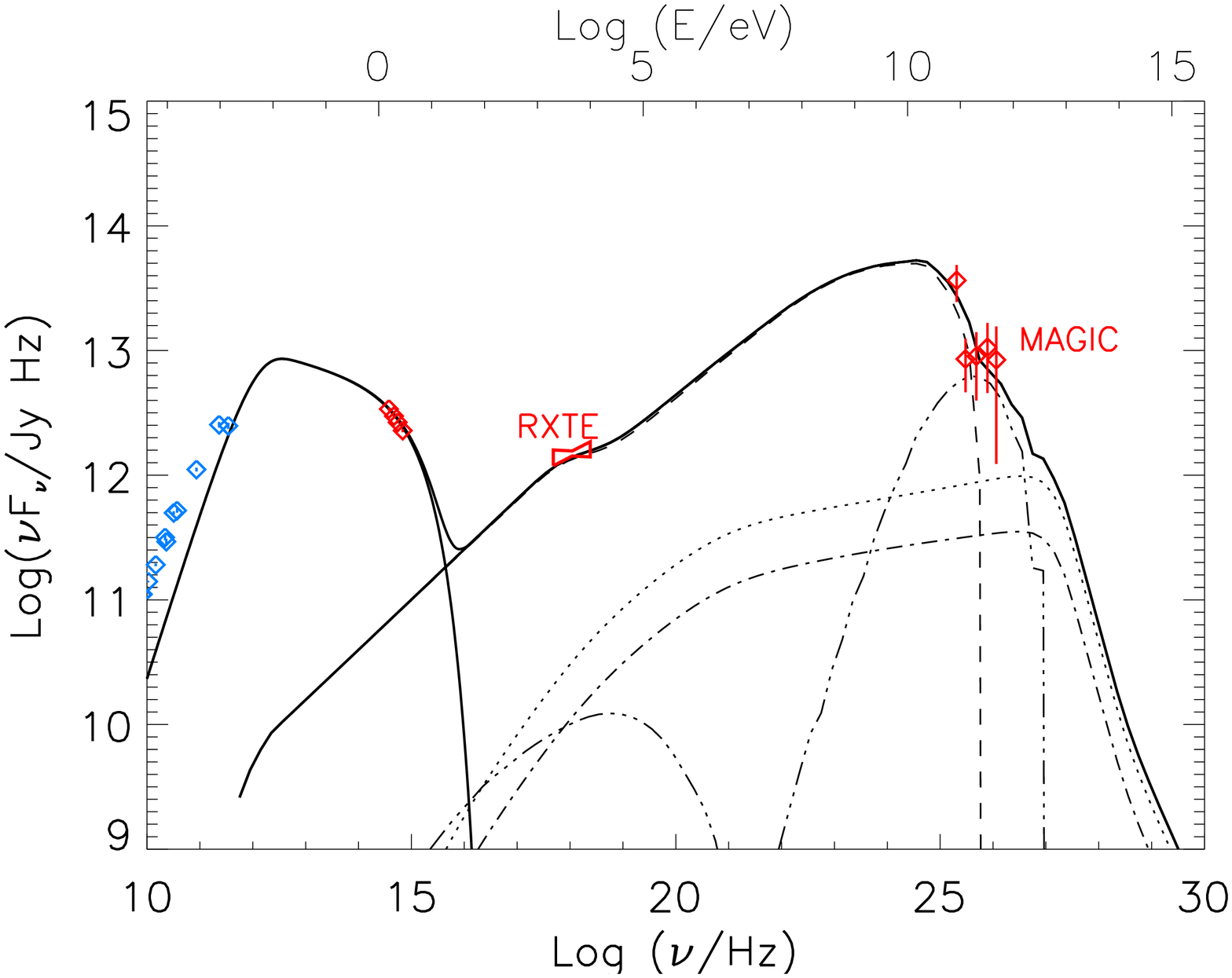}{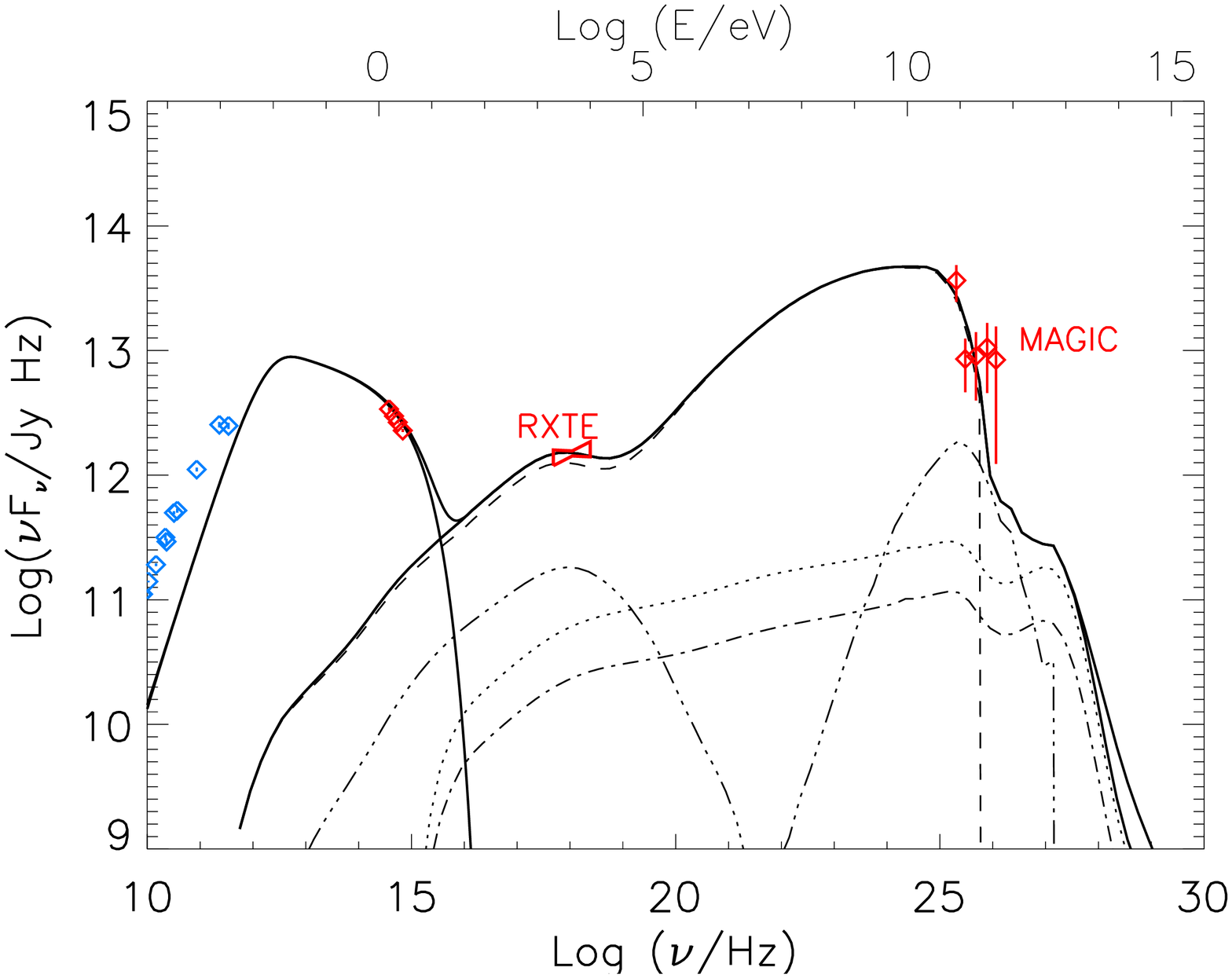}
\caption{Hadronic fits to the SED of 3C279. Left: Pure synchrotron-proton
blazar model with only intrinsic synchrotron photons as targets for
p$\gamma$ pion production; right: synchrotron proton blazar model
including external photon field from the BLR. See table \ref{SPBparameters} 
for parameters. Individual radiation components are: dashed: proton
synchrotron and cascade; triple-dot-dashed: $\mu$ synchrotron and
cascade; dotted: $\pi^0$ cascade; dot-dashed: $\pi^{\pm}$ cascade.}
\label{hfit}
\end{figure}

In the framework of the hadronic SPB model the injection electron 
spectrum is primarily constrained by the optical and radio data. For 
$q_e = 2.1$ -- $2.2$ the primary electron synchrotron spectrum
above the synchrotron self-absorption turnover shows a rather flat 
spectrum from the injected particle spectrum, modified by synchrotron 
losses, followed by a steep tail due to the cutoff of the electron 
distribution at particle Lorentz factors $\sim 10^3$. The high energy 
component of the SED of 3C~279 is constrained by the RXTE and MAGIC data.
In the SPB model the RXTE data can in general be explained by either 
proton synchrotron radiation, or a reprocessed/cascade component. The 
former suggests, however, extremely long loss time scales (of order years 
for typical field strengths and Doppler factors in the SPB model), which 
is difficult to reconcile with the observed day-scale variability. We 
therefore concentrate on the second option of reprocessed radiation dominating
the X-ray band. This picture is also strengthened by the softness of the 
X-ray spectrum, which may indicate the appearance of reprocessed 
(through $\gamma\gamma$ pair production) radiation in this energy range.

The left panel in Fig.~\ref{hfit} shows a typical model fit to the SED 
of 3C~279 of February 23, 2006, for the case of a negligible external
photon field at the location of the $\gamma$-ray emission region. While 
radiation in the $\gamma$-ray band is dominated by proton and $\mu$ 
synchrotron emission, reprocessed proton synchrotron radiation
determines the photon emission in the X-ray band. For the required bulk 
Doppler factors $D = 10$ -- $14$, and assuming a size of the emission 
region of order $\sim 10^{16}$~cm, the energy density of the internal jet 
target photon field amounts to ${u'}_{\rm sy} \sim$ a few $10^{10}$ -- 
$10^{11}$~eV~cm$^{-3}$. With these values, a delay between the TeV and 
X-ray band of a few days may be explainable. The strong magnetic field 
strengths of 40 -- 60~G imply losses due to proton synchrotron
radiation in the TeV band on hour time scales. Interestingly, all 
models representing the simultaneous data of February 23, 2006, reasonably 
well require a cutoff of the injected proton spectrum at a few $10^9$~GeV, 
which is significantly lower than what is needed for HBL-like TeV-blazars. 
The injected proton energy density ${u'}_p$ is of the order $10^2-10^3$~erg~cm$^{-3}$,
not too far from the equipartition value $u'_{\rm p, equi}$ (here ${u'}_p \approx 2$
 -- $17 u'_{\rm p, equi}$), and a total jet luminosity (as measured in the galaxy 
rest frame) of order $10^{48-49}$~erg~s$^{-1}$.

If the gamma-ray emission region is rather close to the BLR, the external 
photon field has to be taken into account as an additional target for 
particle-photon interactions and pair cascading. The observed line and 
disk luminosity of 3C~279 implies $\tau_{BLR} \equiv L_{BLR}/L_D
\approx 0.08$. The additional target photon field in the optical/UV band
leads to enhanced reprocessing in the hadronic model, further softening 
the spectrum in the X-ray band. Signatures of $\gamma$-ray absorption 
due to the additional narrow-banded external photon field are clearly 
visible in the MAGIC energy range. The right panel in Fig. \ref{hfit} 
shows an example of a model fit representing again the February 23, 2006, 
data. Proton synchrotron radiation dominates in the $\gamma$-ray band. 
The reprocessed component is more pronounced here than for the 
case of internal target photon fields only, which yields a more appropriate
description of a potentially steep X-ray spectrum. $\gamma\gamma$
absorption of VHE $\gamma$-ray photons in the external radiation field
inevitably leads to a steepening of the VHE $\gamma$-ray spectrum. The
fit shown in Fig. \ref{hfit}b presents an acceptable representation of
the MAGIC spectrum. Should a higher EBL level imply a harder intrinsic
VHE $\gamma$-ray spectrum, this fit could be modified by choosing a lower
energy density of the external radiation.
Parameter values for 
hadronic model fits are given in Table \ref{SPBparameters}. For the
fit with external photons, the injected proton energy densities are at 
most a factor 4 above the equipartition value and the required total jet 
luminosity of $\sim (1$ -- $7) \times 10^{48}$~erg~s$^{-1}$ is somewhat 
lower than for the case of internal target photon fields only.

\section{\label{summary}Summary}

We have presented simultaneous optical and X-ray spectral information
to the recent MAGIC detection of the quasar-type blazar 3C279 in
February 2006. The source was shown to be in an elevated optical
state, but showed no substantial optical variability and a rather
steep optical spectrum during the MAGIC detection. The GeV -- TeV
flare preceded an X-ray flare by about 5 -- 7 days. We have presented
the simultaneous broadband (optical - X-ray - VHE $\gamma$-rays)
SED of 3C279 and discuss its implication for one-zone leptonic
jet models. We found that an SSC model is extremely problematic
as it would imply unreasonably low magnetic fields. Also an 
external-Compton interpretation has problems with an unusually 
low magnetic field of $B \sim 0.03$~G, implying an equipartition 
ratio of $e_B \sim 10^{-8}$. Alternatively, approximate 
equipartition can be achieved with a bulk Lorentz factor 
of $\Gamma \sim 140 \, R_{16}^{-3/7}$, which appears equally 
unlikely. These constraints have been inferred for a correction of the 
VHE spectrum for $\gamma\gamma$ absorption by the lowest plausible 
EBL model and without taking into account internal $\gamma\gamma$
absorption by photons from the accretion disk or the broad-line
region. Any higher EBL level or a substantial amount of internal
$\gamma\gamma$ absorption would lead to even more extreme
constraints.

We therefore conclude that a simple homogeneous, one-zone leptonic 
jet model has serious problems reproducing the SED of 3C279 on 
February 23, 2006, which includes the recent VHE $\gamma$-ray 
detection by MAGIC. The lack of correlated optical -- $\gamma$-ray
variability suggests, instead, a multi-zone model in which the
optical and $\gamma$-ray fluxes are produced in separate regions
along the jet. We have shown that a leptonic SSC fit to the X-ray
--- VHE $\gamma$-ray spectrum alone can be achieved with parameters 
quite typical for quasar-type blazars. 

Alternatively, the hadronic synchrotron-proton blazar model is
able to provide an acceptable fit to the SED of 3C279. Both a pure
synchrotron-proton blazar model (without external photons as targets
for p$\gamma$ pion production) and a model with a substantial contribution
from external target photons can reproduce the observed SED up to
VHE $\gamma$-ray energies very well. If no external photon field is
included in the model, the relevant proton synchrotron energy loss 
time scale is of the order of years and would therefore be inconsistent 
with the observed day-scale variability if proton synchrotron radiation
was dominant in the X-ray regime. This model therefore requires a high
internal radiation energy density in order for proton-synchrotron
induced cascades to dominate the X-ray emission. Alternatively, an 
additional target photon contribution from external sources (in particular, 
the BLR) is able to overcome this problem, and such a model seems to 
provide an appropriate fit to the observed SED of 3C279 from optical 
to VHE $\gamma$-rays. However, both versions of the hadronic synchrotron
proton blazar model require a rather extreme jet power of 
$L_j \sim 10^{49}$~erg~s$^{-1}$. For comparison, a multi-zone leptonic
model requires a jet power in leptons alone of $L_{\rm j, e} \sim 2.2 \times 
10^{47}$~erg~s$^{-1}$. When including an equal number of cold protons,
the total jet power requirement increases to $L_{\rm j, tot} \sim 2.6 \times
10^{47}$~erg~s$^{-1}$. 

We point out that our conclusions are based on rather incomplete
frequency coverage during the MAGIC detection, and rely on the apparent 
similarity of the SED, at least from radio through X-rays, with previously 
observed high states of 3C279 during the EGRET era. Future simultaneous 
Fermi LAT and TeV $\gamma$-ray observations (with a detection at $> 100$~GeV 
$\gamma$-rays) would provide crucial tests to our conjectures put forth in 
this paper.

\acknowledgments
The work of M. B\"ottcher and was partially supported by 
NASA through XMM-Newton GO grant awards NNX07AR88G and 
NNX08AD67G. The work of A. Marscher was funded in part by
the National Science Foundation through grant AST-0406865 and by NASA
through RXTE Guest Investigator grant NNX06AG86G and Astrophysical Data
Analysis Program grant NNX08AJ64G.

\end{document}